\documentstyle[aps,preprint]{revtex}
\input {amssym.def}
\input {amssym.tex}
\everymath{\rm}
\everydisplay{\rm}

\begin{document}
\title{Electronic and Magnetic States in the Giant Magneto-resistive Compounds}
\author{C. M. Varma}
\address{AT\&T Bell Laboratories \\
Murray Hill, NJ  07974}
\maketitle

\begin{abstract}
The paramagnetic insulator to ferromagnetic metal transition
in Lanthanum manganites and the associated magnetoresistive
phenomena is treated by considering the localization due to
random hopping induced by slowly fluctuating
spin configurations and electron-electron interactions.
The transition temperature and its variation with composition is
derived.
The primary effect of the magnetic field on transport is to alter the
localization length, an effect which is enhanced as the magnetic susceptibility
increases.  Expressions for the conductivity, its variation with magnetic field,
and its connection with magnetic susceptibility in the paramagnetic phase are given
and can be tested with further experiments.
\end{abstract}

\newpage
\section{Introduction}
Interest in mixed-valent compounds of the form
$A_{1-x}^{3+} B_x^{2+} Mn_x^{3+} Mn_{1-x}^{4+} O_3$,
whose fascinating properties were discovered$^1$ about
fifty year ago, has revived recently.
The phase diagram as a function of $x$ for
various trivalent atoms A and divalent atoms B
and the magneto-transport properties has
been determined.$^{2,3}$
For $0.1 \lesssim  x  \lesssim 0.3$ an insulating or very
high resistance phase at high temperatures is followed
at lower temperatures with a metallic phase accompanied by
ferromagnetic order.
The basic physical point,$^4$  made by Zener in 1951, is
that the elementary electron conduction process,
given that valence states other than $3^+$ and $4^+$
are disallowed due to strong local correlations,
in which two Mn ions of different valence
interchange their valence states:
$Mn^{3+} Mn^{4+} \rightarrow Mn^{4+} Mn^{3+}$, is
proportional to the one-electron transfer integral $t$ only
when the initial and final states are degenerate.
This requires that the Hund's rule coupled spins of
$Mn^{3+} ( {\bf S} = 3/2 )$ and of
$Mn^{4+} ({\bf S} = 2)$
are aligned.
Otherwise, the transfer integral is of
$O(t^2 / J)$ where $J$ is the Hund's
rule coupling energy, which for
$Mn^{3+}$ is several eV and assumed much larger than $t$.
An appropriate model Hamiltonian is
\begin{equation}
H = t \sum_{\langle ij \rangle}^{\prime}
c_{i\alpha}^+ c_{j\alpha} +
J \; \sum_i {\bf S}_i \cdot c_i^+
\mbox{\boldmath $\sigma$} c_i
- \mu \sum_i n_i
\end{equation}
where the electrons hop only between nearest neighbor
sites of different valences.
For $| J | \gg t$ the spin of the `conduction' electron
is always parallel to the {\bf local} $S_i$.
Projecting to such states
\begin{equation}
H_{eff} = \sum_{i,j} t ({\bf S}_i  ,
({\bf S}_j + \mbox{\boldmath $\sigma$} )_{max} )
c_i^+ c_j ,
\end{equation}
i.e. the transfer integral  is a function of the relative orientation
of the $Mn^{4+}$ $(S_i =3/2)$
spin and $Mn^{3+} (S=2)$, i.e. ($S_j+ \sigma )_{max}$.

In a pair-wise hopping process the conserved value
of the spin is
$| {\bf S}_i + {\bf S}_j + \mbox{\boldmath $\sigma$} |$,
so we expect the hopping to depend on this function.
Semiclassically the angle $\theta$ between
two spins ${\bf S}_1$ and ${\bf S}_2$ is given by
\begin{equation}
cos \; \frac{\theta}{2} =
\frac{| {\bf S}_1 + {\bf S}_2|}{2S} \; ,
\end{equation}
so that semiclassically the effective hopping Hamiltonian is
\begin{equation}
H_{eff} = \sum_{(i,j)}^{\prime}
t \; cos \frac{\theta_{ij}}{2}
c_i^+ c_j .
\end{equation}
This result was first derived by Anderson and
Hasegawa.$^5$
If one considers pairwise hopping only, one can
specify the relative orientation by $\theta_{ij}$
alone.
More generally,$^6$  one must introduce also the azimuthal
angle $\phi_{ij}$
to specify the relative orientations of spins at
$i$ and $j$
so that a more appropriate form is
\begin{equation}
H_{eff} = \sum_{(i,j)}^{\prime} t \;
e^{i\phi_{ij}}
cos \frac{\theta_{ij}}{2}
c_i^+ c_j .
\end{equation}
$\phi_{ij}$ is a Berry phase.
This may have some interesting consequences which we hope
to discuss in the future.

Since ferromagnetism is accompanied in these materials by
metallicity, one may expect that magnetic polarization
by an external field will cause a large drop in resistivity.
The discovery$^{2,3}$
of such a large magneto-resistance has led to much
interest in these materials.

In the next section, I describe how an adiabatic approximation may be introduced by
first freezing the spin configuration and calculating the electronic states.  It is
argued that the chemical potential passes from a region of localized states to
delocalized states as $x$ is increased.  Spin polaron effects as correction to the
adiabatic approximation are considered next as are electron-electron interaction
effects which are expected to produce a gap in the excitation spectra and in
one-electron states near the chemical potential.  In Section III I calculate the
variation of the ferromagnetic transition temperature with $x$.

\section{General Considerations}
I am primarily interested in the paramagnetic
regime where the spins $\{ S_i \}$ are randomly
oriented and fluctuate at frequencies related only
to the temperature $k_BT$.
Any spin-spin correlation only reduces the
characteristic frequency.
Further assume that $k_BT \ll t$.
Then the core-spin fluctuations may be treated
in the adiabatic approximation, in which
we first freeze the spins in a random configuration
$\{ {\bf S}_i \}$ and calculate the electronic
states
$\psi_n ( \{ S_i \} )$
in the random configuration.
We can then perturbatively couple the thermal
fluctuations of spins to find corrections to the
electronic wave functions.
Simultaneously the potential energy of the spin
configurations is given by
$\sum_n E_n ( \{ S_i \} ) f (E_n)$
where $f$ is the Fermi function.
This procedure is quite analogous to the
Born-Oppenheimer
approximation in electron-phonon problems, except that
no independent inertia exists for the spins.
One can also investigate the paramagnetic
to ferromagnetic transition
by this method.

Consider then first a random configuration of spins.
The transfer integrals for conduction electrons are
then random variables varying from
0 to $|t|$.
The distribution of the transfer integral will be
derived in Section III below.
This randomness was treated recently$^7$ in a
dynamic generalization$^8$ of the coherent
potential approximation (CPA) to calculate the
equilibrium and
transport properties.
In another paper$^9$ it was treated perturbatively
to calculate the resistivity with the conclusion
that the insulating behavior in the paramagnetic
phase is inexplicable in purely electronic models
and suggesting electron-phonon interactions.
This line of thinking has led to the suggestion$^{10}$
that a Jahn-Teller distortion is responsible for the behavior.
The phenomena observed in the manganites occurs also in
mixed-valent $TmSe_xTe_{1-x}$ compounds.$^{11}$
No question of Jahn-Teller distortion arises there.

$LaMnO_3$ is an insulator in which the $Mn-O_6$ octahedra
is Jahn-Teller distorted.
I think it is not correct to conclude that the insulating behavior is
\it due \rm to the Jahn-Teller distortion.
Most transition metal perovskite compounds are insulators, even
$LaTiO_3$ where effective electron-electron interactions on Ti
are smaller than in Mn due to more effective S-electron
screening.
These are all Mott insulators.
Note that the derivation of the double exchange interaction assumes correctly that the direct
electron-electron interaction is even larger than the exchange interaction J.
If symmetry allows a Jahn-Teller distortion in the insulating phase, such
a distortion parasitically does occur.
The conditions for Jahn-Teller effects in metals are much
more stringent and in any case not required by symmetry
and not expected to lead to insulating phase when the metallic bandwidth is $O (2 eV)$.
One expects that the Jahn-Teller distortion of $LaMnO_3$ 
will appear in reduced magnitudes even on addition of Sr in the insulating phase and decrease in 
the low temperature
metallic phase.  More important than the Jahn-Teller distortions for quantitative
purposes are the breathing mode polaronic distortions which are different around
$Mn^{3+}$ and $Mn^{4+}$ because of the large difference in ionic size.

In a model for random hopping, one expects
electronic states to be localized at least
for some energies.
CPA is a (self-consistent) single-site approximation
while localization arises from interference
among scatters.
While the CPA is a
very good approximation for many purposes, it cannot
capture the physics of localization.
I believe the transport properties in this
paramagnetic regime for most of the range of composition as well
as the effect of magnetic field on them can be
understood only when the localization of the single
particle states is considered.

The present problem is one of off-diagonal randomness as
opposed to the Anderson model$^{12}$ for disorder.
This has not been so thoroughly investigated.
One approximate but reliable calculation of the nature
of the wave functions in models of off-diagonal
randomness is due to Economu and Antoniu.$^{13}$
Their results are sketched in fig. (1).
For a model in which the hopping matrix elements $t_{ij}$
are randomly distributed over a 
semi-circular distribution with mean $t_0$ and edges at
$t_0 \mp t_1$
they find that for
$t_1 \approx t_0$, which is relevant for our case,
states in the energy region
$\frac{1}{2} |W| \lesssim E \lesssim |W|$
are localized while these in the middle are extended.
$W$ is the bandwidth of the
bulk which CPA gives correctly as about 0.7 of the total bandwidth.

The localization length in a problem of off-diagonal
disorder is expected to diverge both at the band-edges
and at the mobility edges.
The latter is familiar, and as in models of
diagonal disorder.
The former arises
because states at the band-edges are the exponentially
rare states which travel through the crystal
through routes with identical hopping.

Unlike Ref. (13) the distribution of disorder in the present
problem is not symmetrical.  It follows from eq. (14) below
that the distribution $P (t)$ increases linearly with $t$
with a cutoff at the maximum value.  The higher moments
of the distribution are therefore larger than in Ref. (13).
So the region of localized states is expected to be larger
than in Fig. (1).  Moreover strong correlations in the present
problem (in the presence of disorder) further favor localization.
The effective fluctuations in $t$ are further increased by different polaronic
renormalizations of $t$ and a $Mn^{3+}$ ion and an $Mn^{4+}$ ion.

We may safely assume that in the frozen
spin approximation, states at the chemical
potential are localized for low dopings and that
a mobility edge occurs so that for higher dopings
the states at the chemical potential are delocalized
although strongly scattering.
CPA should be a good approximation to calculate
properties in the second regime.
\subsection*{Spin-polaron Formation: First Correction to the Adiabatic Approximation}
The spins do not have any inertia.
The leading correction to the adiabatic approximation
then amounts to (i) minimizing the free-energy $F(\{ S_i \})$
with respect to spin
configurations $\{ S_i \}$ and (ii) calculating
scattering between electronic states due to fluctuations
about the new configuration.
The first part is the same as considering spin-polaron
formation.

We can easily make the first correction to the
frozen random spin-configuration by the formation
of spin polaron around an electronic state which
otherwise would be highly localized.
Let there be $P$ lattice sites which are spin-polarized
so that an electron can hop freely between them to lower
its kinetic energy.
This is opposed by the entropy lost by the spins.
So $P$ is determined by minimizing
\begin{equation}
E(P) = -W +
\frac{at_0}{P^{2/3}} +
PkT \; ln \; (2S+1)
\end{equation}
where $a$ is a numerical factor of O(1),
so that
\begin{equation}
P \approx (at_0 /T)^{3/5}
\end{equation}
This is an elementary generalization to finite temperature
of Nagaoka's result$^{14}$ for such models that, at $T=0$, a
single carrier will induce ferromagnetic order.
In the manganites the effective $t$ (renormalized for effect
other than spin fluctuations, phonons for instance)
is about 0.2 eV.
So at room temperatures the size of the spin-polaron
is only a few bonds.
So the spin-polaron effect appears not to invalidate the adiabatic approximations
over the bulk of the band in the paramagnetic range of temperatures of interest.

A prediction following from these considerations is that
the effective moment in susceptibility measurements
even at fairly high temperatures will be larger than that of
the appropriate average of $Mn^{3+}$ and $Mn^{4+}$
moments.
At a concentration $x$ of the spin
$S_1$ and the rest $S_2$,
one expects
\begin{equation}
S_{eff}^2 = x (S_1+PS_2)
(S_1+PS_2+1) +
(1-x-Px) \; S_2 (S_2+1)
\end{equation}
The estimate of spin-polaron effects would
be quite altered for states in the tails of Fig. (1) or those
near the mobility edge which have a large localization
length.
The polaron effects would have a major effect only on the surface
of such states leaving the bulk of it relatively
unaffected.
So Eq. (8) is an overestimate.
Moreover (8) is merely the first sign of fluctuations
above $T_c$ which increase the susceptibility over the
Curie-law.

The adiabatic assumption made here is accompanied
by the ergodic hypothesis.
The spin-configurations change locally changing
the localization length of the electronic state
in a given small region.
But averaged over the sample the distribution of the energies
and localization lengths of the states stays the same.
This process does not invalidate the adiabatic
approximation unless the temperature is so high
that several states of energy $\lesssim 0(kT)$
larger than the chemical potential have spatial
overlap with states at the chemical potential.
This stability argument gets stronger due to the role
of electron-electron interactions in depleting states
near the chemical potential discussed below.

The last paragraph is only a plausibility argument.
It is a very interesting unsolved theoretical
problem to ask for the frequency scale up to
which disorder can vary without delocalizing states.
\subsection*{Electron-electron Interactions}
The role of electron-electron interactions is always
more important for localized states where the
kinetic energy has been quenched than for Bloch states.
Efros and Shklovskii$^{15}$ have given convincing arguments
that Coulomb interactions create a pseudo-gap at the
chemical potential if it lies below the mobility
edge.
Therefore due to electron-electron interactions the density
of {\bf one-particle states} is modified from
Fig. (1) to that depicted in Fig. (2).
In three dimensions, the density of states in the
localized regime per unit volume near the
chemical potential is of the form
\begin{equation}
\begin{array}{rl}
g( \epsilon ) & = \epsilon^2 / \Delta^3 \\
\Delta & = \alpha \frac{e^2}{\kappa}
\end{array}
\end{equation}
where $\alpha$ is a numerical constant and
$\kappa$ is the dielectric constant.
$g( \epsilon )$ is independent of the localization length near
the chemical potential.
Equation (9) is arrived
at by considering stability of one particle
excitations under excitonic renormalizations
due to the Coulomb interactions.
Corrections to it due to multiparticle
excitations have not yet been established
conclusively.

The best way to experimentally test the localization idea
and (9) is a tunneling measurement.
Density of states of the form (9) have however already
been observed in photoemission experiments.$^{16,17}$
The key point is that the density of states
is zero at the chemical potential (with $T/ \Delta$
corrections) independent of doping if the chemical
potential is below the mobility edge and the
experiment is done in the paramagnetic regime.

\subsection*{Resistivity}
The conductivity of the localized states
is expected to be of the variable range
hopping form:
\begin{equation}
\sigma \sim \; exp \; [-(T_0 /T)^{1/2} ]
\end{equation}
with $T_0 \approx e^2 / \kappa \ell$,
for $T \ll T_0$.
Here $\ell$ is the localization length for states
near the chemical potential.
As the effective disorder is decreased by
applying a magnetic field, $\ell$ increases leading
to a decrease in the resistivity.
The relationship of the effective disorder and the magnetization is
derived in the next section.
In the paramagnetic regime, the leading dependence of the
localization length on $H$ is
\begin{equation}
\ell (H) = \ell_0 (1+ \chi (T)
H^2 / \bar{t} ) ,
\end{equation}
where $\ell_0$ depends on the electron density and
$\bar{t}$ is $O(t)$.
Equation (10) may be combined with Eq. (11) to suggest that
magnetoresistance plotted against the magnetic susceptibility
should fall on the same curve for different temperatures and for
different compositions.
Similar scaling behavior should be observed for other transport
properties as well.

Note that the leading temperature dependence of the
conductivity (or magneto-conductivity) exhibited in Eq. (10)
gives the number of carriers participating in the conduction.
Their mobility is a weakly temperature (and field) dependent
pre-factor of (10).
This is consistent with the recent observations of Ong et al.$^{18}$
from Hall effect and magnetoresistance measurements that the
dominant effect of magnetic fields is to increase the number
of carriers.

The frequency dependent conductivity should exhibit the effects of the
Efros-Shklovskii gap in the paramagnetic phase.
A gap is indeed observed.$^{19}$
The low energy $\sigma ( \omega )$ appears roughly $\sim \omega$.
This is quite remarkable and needs more detailed consideration.

When $T_0 \sim T$, the conductivity is no longer of the
form (10).
Now several excited states of energy $\leq 0 (kT)$ overlap
the states at the chemical potential.
The mobility may now be calculated from
\begin{equation}
\mu = \frac{eD}{kT}
\end{equation}
where the diffusion constant
$D \approx \ell^2 / \tau$; $\ell$ is the
typical distance between states degenerate to within
$kT$ which at very high temperatures is $0(a)$
the lattice constant; and $1/ \tau$ is the typical
frequency of spin fluctuation $\sim 0 (kT)$.
So the high temperature conductivity
approaches a constant proportional to the carrier density.
\section{Estimate of Metallic/Ferromagnetic Transition Temperature}
We can calculate the equilibrium properties in the
paramagnetic phase and an estimate of the ferromagnetic
transition temperature $T_c$ by a variational calculation
of the free-energy.
Let the probability distribution of the angle between
neighboring spins $\theta$ be given by
\begin{equation}
P(x) = M^2 \delta (1-x) +
(1-M^2) {\cal N} \; exp
\left [ - \frac{A}{kT} (1-x ) \right ]
\end{equation}
where $x= cos \; \theta$, $M$ is the uniform magnetization
which is finite for $T>T_c$ only if a magnetic field $H$
is applied, A is a variational parameter which is a
function of temperature and ${\cal N}$ is a normalization factor.
$P(cos \theta )$ for $A/kT \rightarrow \infty$
aligns all the spins while for $A/kT \rightarrow 0$,
the spins are completely randomly oriented provided
$M^2 =0$.
So it is expected that as temperature decreases
the variational calculation will lead to
$A(T)/kT$ changing from 0 to very large
values near the ferromagnetic transition.
To calculate the energy of the electrons, we evaluate
the probability distribution of the nearest neighbor
transfer integral $P(t)$:
\begin{equation}
\begin{array}{rl}
P(t) &= \frac{1}{t_0}
\int_{-1}^1 d(cos \theta )
P ( cos \theta ) \;
\delta (t/t_0 - cos \theta /2 ) \\
&= M^2 \delta (t-t_0) +
(1-M^2) {\cal N} \frac{t}{t_0^2}
exp \left [ - \frac{2A}{kT}
\left ( \frac{t^2}{t_0^2}-1 \right ) \right ]
\end{array}
\end{equation}
One should now calculate the electronic density
of states $n(E)$ with $P(t)$ and calculate
the electronic energy in terms of $n(E)$.
$n(E)$ can not be evaluated without a lengthy
numerical calculation.
We will instead use the following approximate
method:
For a rectangular density of states with bandwidth $W$,
with height $W^{-1}$ so that it can accommodate at
most one electron per atom, the energy at $T=0$ for
$c$ electrons per atom is
\begin{equation}
\epsilon (t) = - \frac{W}{2}
c(1-c) \; .
\end{equation}
A good approximation for the electronic
energy is
\begin{equation}
E_e = \int dt \; P(t) \epsilon (t) .
\end{equation}
Classical approximation to the entropy of
spins in the large $S$ limit using
distribution such as $P(x)$ give incorrect
answers at high temperatures.
To calculate the entropy we write the quantum version of
$P(x)$ by writing
\begin{equation}
cos \theta -1 =
\frac{J^2-4S^2}{2S^2} ,
\end{equation}
where $S$ is the value of the spin of an ion
($Mn^{3+}$ spin $S=2$ is the appropriate value) and $J$
ranges from 0 to $2S$ in integer increments.
Then define the quantum analog of the
second tern of (1)
\begin{equation}
Z = \sum_J e^{-\frac{A}{kT} S^2 (J^2-4S^2)}
\end{equation}
and
\begin{equation}
S_{ions} = - \frac{\partial}{\partial T}
(kT ln Z)
\end{equation}

Strictly speaking I should use (18) for the calculation of the energy
also,
but the final answer is likely to be nearly the
same and the form of $P(x)$, equation (13) leads to
$P(t)$ of Eq. (14) which seems physically more transparent.

In my calculations I have neglected the entropy of the
orbital motion of the $x$/unit cell ``conduction''
electrons compared to that of the 1/unit cell spins.
This is a small correction.
Similarly finite temperature correction to (3) were neglected.

Free energy then is
\begin{equation}
F \approx \; \; E_e - T \; S_{spins} - {\bf M \cdot H}
\end{equation}
We may write this as
\begin{equation}
F = t_0 [ \alpha (A , \; T/t_0)
{\bf M}^2 + F_0 (A^2 , T/t_0) ] - {\bf M \cdot H},
\end{equation}
so that the uniform magnetic susceptibility is given by
\begin{equation}
\chi^{-1} = 2 t_0 \alpha
(A, T/t_0)
\end{equation}
The ferromagnetic transition temperature is given by
\begin{equation}
\alpha (A^2 , T_c/t_0) =0
\end{equation}
The result of the minimization of the free-energy
with respect to A, not surprisingly (since this is a
mean-field calculation), is that $\chi (T)$
follows a Curie law
\begin{equation}
\chi (T) \approx \frac{S^2}{T-T_c}
\end{equation}
The numerical minimization of F with respect to A gives that
\begin{equation}
k_BT_c \simeq 0.1 \; E_{coh}^F (c)
\end{equation}
where $E_{coh}^F (c)$ is the electronic cohesive energy
for the ferromagnetic case
\begin{equation}
E_{coh}^F (x) =
\frac{W}{2} c (1-c) .
\end{equation}
The numerical factor of $\sim 0.1$ reflects (i) that
the electronic cohesive energy increases by only
about 20\% in going from the complete random spin-orientation
to the ferromagnetic configuration and (ii) that the
entropic free energy of spins at a temperature
$T=W$ is about a factor of 2 larger than $W$.

Mattheiss has calculated the conduction electron
bandwidth in the local density approximation to be
$\approx 2.5$ eV.
Then at $x \approx 0.3$, the calculated
$T_c \approx 250$ which, considering the crudeness
of the calculation, is in the right range.
A more important test is the relation
$T_c \sim x(1-x)$ which is compared with experimental
results in Fig. (2).

CPA is a perfectly respectable approximation to
calculate the energetics.
Therefore Furukawa's calculations$^7$ for $T_c$ are probably 
better than the estimates here although the
considerations here may be
more transparent.
\section{CONCLUDING REMARKS}
An attempt has been made to understand the
equilibrium and transport properties of mixed-valent
magnetic compounds on the basis of electronic localization
due to magnetic disorder and the alteration of this disorder by a
magnetic field.
A paramagnetic insulator to Ferromagnetic transition in the observed range of temperatures
and with the observed composition dependence is derived.  The primary effect of a magnetic
field on transport is to alter the localization length, an effect which is enhanced as the
magnetic susceptibility increases.  Scaling of the transport properties with the magnetic
susceptibility and tunneling experiments to observe the Efros-Shkloveskii correlation
induced pseduo-gap in the paramagnetic phase are suggested.

In this paper the ordinary antiferromagnetic exchange
between localized carriers and the effects due to A-B
disorder have been ignored.
At low temperatures their effects have been adequately
treated by deGennes.$^{20}$

\subsection*{Note Added}
Based on a preprint of this note, the authors of Ref. (10) have in yet another paper on
the Jahn-Teller distortion idea (A. J. Millis, R. Mueller, B. I. Shraiman, preprint)
pointed out in effect that the fluctuations in hopping estimated here using the results
of Economu and Antoniu are small by about a factor of 2 to give the mobility edge at
$x \approx 0.3$.  This is picayune.  The point of this work is to show an important
effect which systematically explains observations as $x$ is varied.  Estimates of the
fluctuation in hopping based on ``one-electron approximation'' can easily be in error by
a factor of two or more.  As pointed out here electron correlation effects increase the
tendency to localization and increase the effective disorder parameter.  So does the
different lattice distortion (breathing mode) around $Mn^{3+}$ and $Mn^{4+}$ which
renormalizes $t$.  The uncertanity in disorder fluctuations by a factor of 2 should be
compared with a general lack of knowlege of coupling constrants in the Jahn Teller
scheme.  In any case recent neutron scattering results strongly disfavor Jahn Teller
effects as responsible for the insulating to metallic crossover (M. Marezio, private
communications).

\subsection*{Acknowledgements}
I wish to acknowledge useful discussions with G. Aeppli, B. Batlogg,
S. Cheong, H. Hwang, P. B. Littlewood, A. J. Millis, T. Paalstra,
A. Ramirez, B. I. Shraiman, Y. Tokura, and especially H. Mathur and A. Sengupta.
\clearpage
\section*{REFERENCES}
\begin{enumerate}
\item G. H. Jonker and J. H. van Santen,
Physica (Utrecht) {\it 16}, 337
(1950); {\it 19}, 120 (1953).
\item
R. M. Kusters et al., Physics (Amsterdam) {\it 155}B,
362 (1989); K. Chahara et al., Appl. Phys. Lett. {\it 63},
1990 (1993); M. McCormack et al., Appl. Phys. Lett. {\it 64},
3045 (1994); G. C. Xiong et al., Solid State Comm. (to be published);
Phys. Rev. (to be published); H. Y. Hwang et al., Phys. Rev. Lett.
{\it 75}, 914 (1995); J. H. Schriffer et al., Phys. Rev. Lett.
{\it 75}, 3336 (1995).
\item
Y. Tokura et al., J. Phys. Soc. Japan {\it 63},
3931 (1994);
A. Asamitsu et al., Nature (London)
{\it 373}, 407 (1995); S. Jin et al., Science {\it 264},
413 (1994).
\item
T. Paalstra et al., (Preprint);
C. Zener, Phys. Rev. {\it 82}, 403 (1951).
\item
P. W. Anderson and H. Hasegawa,
Phys. Rev. {\it 100}, 675 (1955).
\item
H. Mathur (Private Communication).
\item
N. Furukawa, J. Phys. Soc. Japan {\it 63},
3214 (1994) and Preprints.
\item
A. George, G. Kotliar, M. Rozenberg,
Rev. Mod. Phys. (in print).
\item
A. J. Millis, P. B. Littlewood and
B. I. Shraiman,
Phys. Rev. Lett. {\it 74}, 5144 (1995).
\item
A. J. Millis, B. Shraiman and M. Mueller, (preprint).
\item
C. M. Varma, Solid State Comm, {\it 30}, 537 (1979);
B. Batlogg, H. R. Ott, and P. Wachter, Phys. Rev. Lett.
{\it 42}, 282 (1979).
\item
P. W. Anderson, Phys. Rev. {\it 109},
1492 (1961).
\item
E. N. Economu and P. D. Antoniu,
Sol. State Comm. {\it 21}, 285 (1977).
\item
Y. Nagaoka, Phys. Rev., {\it 147}, 392 (1966). 
\item
B. I. Shklovskii and A. L. Efros,
Chapter 10 in {\it Electronic Properties of
Doped Semiconductors}, Springer-Verlag,
Berlin (1984).
\item
T. Saitoh et al., Phys. Rev. B{\it 51},
13942 (1995), and lecture at
Naeba Conference, Japan, Oct. 31 (1995).
\item
J. H. Park et al., preprint.
\item
N. P. Ong, (Private Communication).
\item
Y. Okimoto, Phys. Rev. Lett. {\it 75}, 109 (1995).
\item
P. G. deGennes, Phys. Rev., {\it 118}, 141 (1960).
\end{enumerate}
\clearpage
\section*{FIGURE CAPTIONS}
\begin{itemize}
\item[Fig. 1a]
Character of states as a function of
their energy in a tight binding
model with disorder in which $t_0$ is the
mean and $t_1$ the rms
width of the transfer integral.
This sketch is based on Economu
and Antoniu, Ref. (13).
\item[Fig. 1b]
Sketch of the density of states.
\item[Fig. 2]
Density of states with Efros-Shklovskii gap.
\item[Fig. 3]
Magnetic transition temperature vs x in
$La_{7-x}Sr_xMnO_3$ taken from Ref. (3),
compared to the result $T_M \sim x(1-x)$
from Eq. (25).
\end{itemize}
\end{document}